\documentclass[floatfix,reprint,amsmath,amssymb,aps,prapplied,
]{revtex4-2}

\usepackage{graphicx}
\usepackage{dcolumn}
\usepackage{bm}
\usepackage{booktabs}
\usepackage{hyperref}
\DeclareUnicodeCharacter{202F}{\,}

\hypersetup{
    colorlinks=true,
    linkcolor=blue,
    filecolor=magenta,
    citecolor=blue,
    urlcolor=blue
}

\newcommand{\ket}[1]{\left|#1\right\rangle}
\newcommand{\bra}[1]{\left\langle#1\right|}
\newcommand{\Rb}[0]{$^{87}$Rb~}

\begin{document}

\preprint{APS/123-QED}

\title{Microwave-optical double-resonance vector magnetometry with warm Rb atoms}

\author{Bahar Babaei}
\author{Benjamin D. Smith}%
\altaffiliation[Current address: ]{Sandia National Laboratories, Albuquerque NM, USA}
\author{Andrei Tretiakov}
\altaffiliation[Current address: ]{Department of Physics, University of California, Los Angeles, Los Angeles, CA, USA}
\author{Andal Narayanan}
\author{Lindsay J. LeBlanc}
 \email{lindsay.leblanc@ualberta.ca}
 \affiliation{
Department of Physics, University of Alberta, Edmonton AB T6G 2E1, Canada
}

\date{\today}

\begin{abstract}
 Non-invasive, accurate vector magnetometers that operate at ambient temperature and are conducive to miniaturization and  self-calibration hold significant potential. Here, we present an unshielded three-axis vector magnetometer whose operation is based on the angle-dependent relative amplitude of magneto-optical double-resonance features in a room-temperature atomic ensemble. Magnetic-field-dependent double-resonance features change the transmission of an optical probe tuned to the D2 optical transition of $^{87}$Rb in the presence of a microwave field driving population between the Zeeman sublevels of the hyperfine ground states $F = 1$ and $F = 2$. Sweeping the microwave frequency over all Zeeman sublevels results in seven double-resonance features, whose amplitudes vary as the orientation of the external static magnetic field ($\vec{B}_{\text{ext}}$) changes with respect to the optical and microwave field polarization directions. 
 Using a convolutional neural network model, the magnetic field direction is measured in this proof-of-concept experiment with an accuracy of $1^{\circ}$ and its amplitude near  50~$\mu$T  with an accuracy of 115~nT. 
\end{abstract}

\maketitle

\section{\label{sec:introduction}Introduction}
As demands upon the accuracy and precision of sensors increase, efforts in both research and industry are increasingly looking towards opportunities with quantum systems, which capitalize on  coherence, superposition, or entanglement to achieve sensitivity levels that outperform classical methods~\cite{Degen2017-it}.
Quantum sensing with alkali atoms has significant potential in  applications where external fields and forces must be measured with extraordinary resolution, enabled by the atoms' relatively simple electronic structure and strong optical couplings~\cite{Simons2021, Cai2020, Behbood2013, Bison2018}. Precision magnetometers are increasingly essential across scientific and technological fields, including biomedical imaging such as magnetoencephalography~\cite{Brookes2022, Aslam2023, Murzin2020, Nardelli2020, Sander2020}, space missions~\cite{Korth2016, Slocum1963,Cochrane2016}, and geophysical surveys~\cite{Nabighian2005, Ingleby2018, Hrvoic2005}.

Atomic magnetometers based on room-temperature alkali vapor have garnered particular attention due to their combination of high sensitivity, self-calibration, and flexibility for commercialization~\cite{Bloom1962, Sebbag2021, Shah2007}. While cold-atom techniques enable higher spatial resolution, the demanding experimental conditions, including vacuum systems and the limited atom number constrain their applicability~\cite{Gawlik2013, Cohen2019,  Fabricant2023}. Ambient-condition atomic magnetometers using hot or room-temperature alkali atom vapors, which operate based on spin-exchange relaxation-free (SERF) magnetometry~\cite{Sheng2013, Seltzer2004, Papoyan2016, Budker2002}, nonlinear magneto-optical rotation (NMOR)~\cite{Patton2014, Pyragius2019, Gawlik2006}, and electromagnetically induced transparency (EIT)~\cite{Stahler2001, Yudin2010, Cox2011,Das2021-qp}, exploit distinct atomic interactions to achieve high sensitivity in different configurations, and may be arranged to  measure both magnetic field amplitude and orientation using magneto-optical effects. These atomic vector magnetometers typically require two or three optical beams, radio frequency (RF) coils, and/or magnetic shielding to extract the full vectorial information of $\vec{B}_{\text{ext}}$ without mechanical rotation~\cite{Brekke2017, Dhombridge2022, Yao2022, Qin2024, Zou2024}.

In this work, we present a three-axis vector magnetometer based on optical absorption in a room-temperature rubidium vapor, based on magneto-optical double-resonance~(DR), which uses a single optical beam and a microwave cavity. This microwave-mediated optical magnetometer leverages the distinct, quantized angular momentum levels of room-temperature $^{87}$Rb atoms, and makes use of atom-field interactions in both the microwave and optical frequency regimes~\cite{Steck2024, Kiehl2025}. 
The orientation of the optical and microwave fields provides two reference axes with respect to which the orientation of $\vec{B}_{\text{ext}}$ can be determined. The DR sensitivity to both the amplitude and orientation of microwave radiation in this atomic system  offers a promising avenue for advancing microwave magnetometry~\cite{Wang2015, Bohi2010,Wang2021-ei}.
Multi-level atomic systems, such as rubidium-87 with its eight ground- and sixteen excited-states,  provide both richness and complexity. One one hand, multiple degrees of freedom can be extracted from a single measurement; on the other, it is increasingly challenging to analyze the interactions and to distill vectorial information of $\vec{B}_{\text{ext}}$. Additionally, processes such as optical pumping and collisional spin-exchange that redistribute atomic populations make modelling the exact conditions difficult. Machine learning provides a unique opportunity to address such challenges in sensing applications that involve limited parameters and complex physics~\cite{ Meng2023, Gonzalez2024}. Here, we demonstrate how a convolutional neural network (CNN) can use measurement data to accurately determine the amplitude and two spatial angles of $\vec{B}_{\text{ext}}$ in the polar coordinates.
\begin{figure*}
    \centering
    \includegraphics{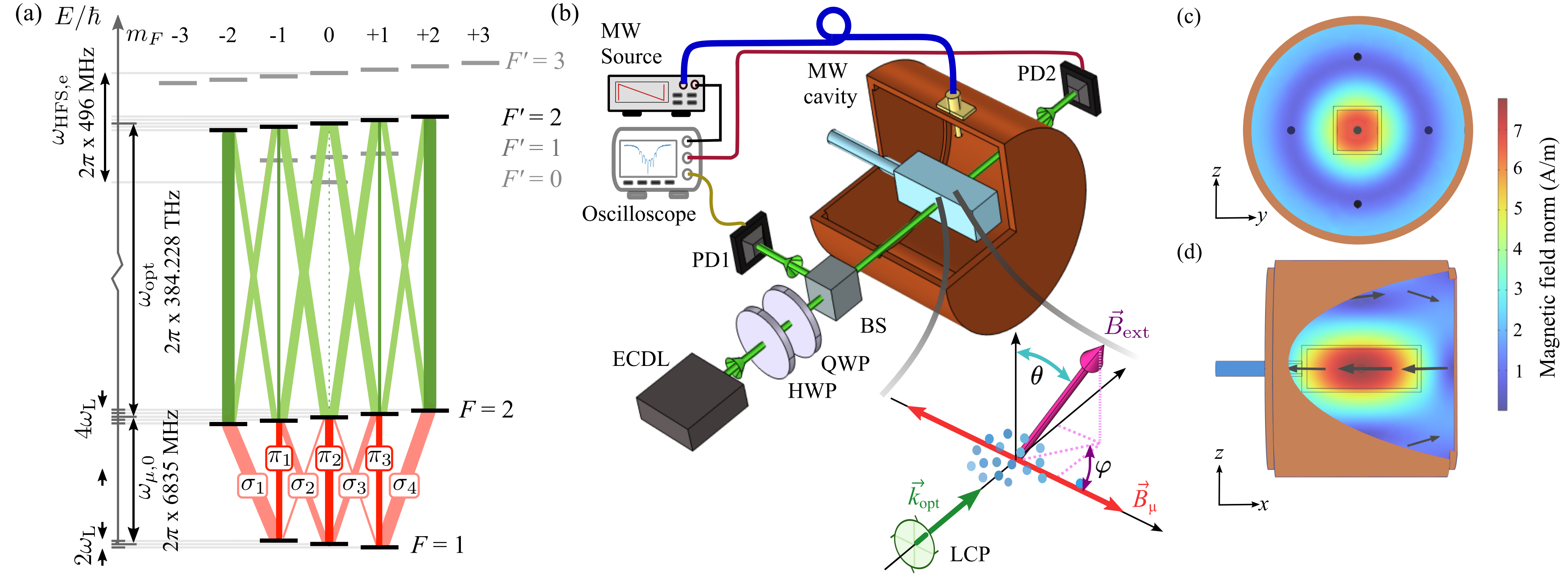}
    \caption{(a) Allowed microwave and optical transitions in \Rb between Zeeman sub-levels are shown as red and green paths. The thickness of these paths is proportional to the branching ratio coefficients, representing the transition coupling strength. Lighter diagonal paths indicate $\sigma$ transitions, while darker vertical paths correspond to $\pi$ transitions. (b) Experimental setup, including a vapor cell, a cylindrical microwave (MW) cavity, an external cavity diode laser (ECDL), half-wave plates (HWP), a quarter-wave plate (QWP), a polarization beam splitter (PBS), a beam splitter (BS), and photodiodes (PD). 
    Inset with coordinates shows $\vec{B}_{\text{ext}}$ (pink arrow),  magnetic component of microwave radiation ($\vec{B}_\mu$) (red arrow), and left-circular polarized (LCP) optical radiation (green arrow). (c) Cross-sectional (top) and side (bottom) views of the  simulated  $\text{TE}_{011}$ mode in the cylindrical microwave cavity, including the vapor cell. The normalized field amplitude  is indicated by color,  and direction of $\vec{B}_\mu$ (at an arbitrary moment) within the microwave cavity by arrows.  The copper color represents the cavity's outer surface. }
    \label{fig:intro}
\end{figure*}
\section{\label{sec:Magneto-optical}Vector magnetometry via magneto-optical double-resonance}

Double-resonance spectroscopy is well-studied as a tool for measuring weak microwave or radio-frequency transitions in atomic ensembles~\cite{Qi2020, Sheng2013, Sheng2017, Weis2006,Tretiakov2019-yd}.  The DR condition is satisfied when two resonant fields are present in a three-level system, typically addressing one optical and one microwave or radio-frequency transition. In scalar DR magnetometry~\cite{Ingleby2018, Bevilacqua2016},  the atomic spin is modulated at the Larmor frequency using the low-frequency field (corresponding to transitions between hyperfine or Zeeman sublevels), and the corresponding modulation in optical absorption is recorded. 

\subsection{Double-resonant absorption}\label{DR absorption} 

In steady-state absorption, an optical probe through a three-level vapor will result in optical pumping, wherein the ground state on the probe transition is depopulated towards the uncoupled ground state; this results in reduced optical absorption and greater probe transmission.  When the ground states are addtionally coupled by a resonant lower-frequency field  $\omega_\mu$ (DR configuration), the first ground state is repopulated, resulting in greater absorption of the probe and reducing the optical transmission. If $\omega_\mu$ is swept across the resonance frequency between the ground levels $\omega_{12}$, a dip in the transmission will be observed. At zero external magnetic field there is a single resonance; polarization-dependent  absorption emerges when the external field is non-zero and lifts the ground state degeneracies.

The full DR energy-level configuration is illustrated in Fig.~\ref{fig:intro}(a), where the presence of $\vec{B}_{\text{ext}}$ breaks the degeneracy in the hyperfine levels. 
To begin to understand how optical absorption is affected by the parameters in this system, we use a simplified model wherein we consider only three levels from this example system.
The ground Zeeman sublevels $\ket{1}\equiv\ket{F=1, m_F}$ and $\ket{2}\equiv\ket{F=2, m'_F}$,  energetically split by $\hbar\omega_{12}$, are connected by the magnetic dipole transition with an electromagnetic field at frequency $\omega_\mu$, while an electric dipole transition (driven by a continuous wave at the optical D2 transition at frequency $\omega_{\rm opt}$) connects ground level $\ket{2}$ and excited level $\ket{e}\equiv\ket{F'=2, m''_F}$, which are split by energy $\hbar\omega_{2e}$.
The strengths of the couplings are quantified by the Rabi frequencies: for the magnetic dipole transition  $\Omega_\mu = -{\langle 2 | \vec{\mu}_s \cdot \vec{B}_\mu(r) | 1 \rangle}/{\hbar}$, where $\vec{\mu}_s$  is the magnetic dipole moment and  $\vec{B}_\mu(r)$ is the the magnetic component of the microwave radiation; for the electric dipole transition,  $\Omega_{\rm opt} = {\langle e | \vec{d}  \cdot \vec{E}_{\rm opt}(r)  | 2 \rangle}/{\hbar}$, where  $\vec{d} $ is the electric dipole moment and $\vec{E}_{\rm opt}(r)$ is the electric component of the optical radiation.

In the usual rotating-wave approximation, the three-level Hamiltonian for the atomic and interaction terms, with the dual microwave and optical interactions is 
\begin{equation}
\hat{\mathcal{H}} = -\hbar
\begin{pmatrix}
0 & \Omega_{\mu} e^{i\omega_{\mu} t} & 0 \\
\Omega_{\mu} e^{-i\omega_{\mu} t} & -\omega_{12} & \Omega_{\rm opt} e^{i\omega_{\rm opt} t} \\
0 & \Omega_{\rm opt} e^{-i\omega_{\rm opt} t} & -(\omega_{2e}+\omega_{12})
\end{pmatrix}.
\end{equation}
To study the effect of the microwave transition on the optical transmission, we consider the density matrix describing the three levels and solve the Lindblad master equation for the steady-state solutions (Appendix~\ref{app:A}). From this, we see that the absorption on the optical transition depends on the relative populations in levels $\ket{2}$ and $\ket{e}$. In the case where  both the optical and microwave fields are resonant, we use the rotating-wave approximation (under the conditions for optical transition linewidth $\Gamma$ and ground state decoherence $\gamma$: $\Gamma \gg \Omega_\mu \gg \gamma$) to approximate the absorption coefficient for the optical transition 
\begin{equation}
A_{\rm DR} \propto  \frac{N L I_0\left\{\gamma\Gamma + \Omega^2_\mu \left[\left(\frac{\Gamma}{\Omega_{\rm opt}}\right)^2 - 2\right]\right\}}{3\gamma\Gamma + 2\Omega^2_{\rm opt} + 2\Omega^2_\mu \left[\left(\frac{\Gamma}{\Omega_{\rm opt}}\right)^2 + 2 \right]},
\label{DRamp}
\end{equation}
where $N$ is the atomic density, $I_0$ is the incident optical intensity (proportional to $\Omega_{\rm opt}^2$), and $L$ is the length through which the absorption beam travels. Equation~\ref{DRamp} reveals the strong dependence of the  absorption on the microwave and optical Rabi frequencies, as well as on the optical depth. All the information about $\vec{B}_{\text{ext}}$ is embedded in these Rabi frequencies, which depend on the field amplitude and polarization experienced by the atoms and on the transition strength between initial and final states. 

While the absorption described by Eq.~\ref{DRamp} is considerably simpler than the experimental conditions described next, the operational principle is revealed: in the DR configuration, optical absorption (and transmission) depends on both fields addressing the levels in question through transition matrix elements that depend on the orientation of the fields' polarization with respect to a quantization direction. For the atomic transitions used here, that quantization direction is determined by the external magnetic field $\vec{B}_{\rm ext}$ to be measured, and thus information about its direction is part of the transmission signal.

\subsection{Dependence of absorption features on the external magnetic field}\label{Absorption dependence on external magnetic field}

  To move beyond scalar magnetometry and determine the full vector information of the magnetic field, we employ a DR technique that addresses microwave and optical transitions in an ambient-temperature \Rb vapor. A cavity-enhanced continuous microwave field~\cite{Tretiakov2020-py,Ruether2021-rq}  addresses the magnetic dipole transition between the $\ket{1}$ and $\ket{2}$ hyperfine levels of the ground state at $\omega_{12}/2\pi = 6.834$ GHz. By scanning the microwave frequency across the multiple resonances of the Zeeman-split $\ket{1} \rightarrow \ket{2}$ transition, and measuring the relative transmission of the optical probe at each, we gain the multiparameter information required to characterize the complete magnetic field vector.

\begin{figure}
    \centering
    \includegraphics{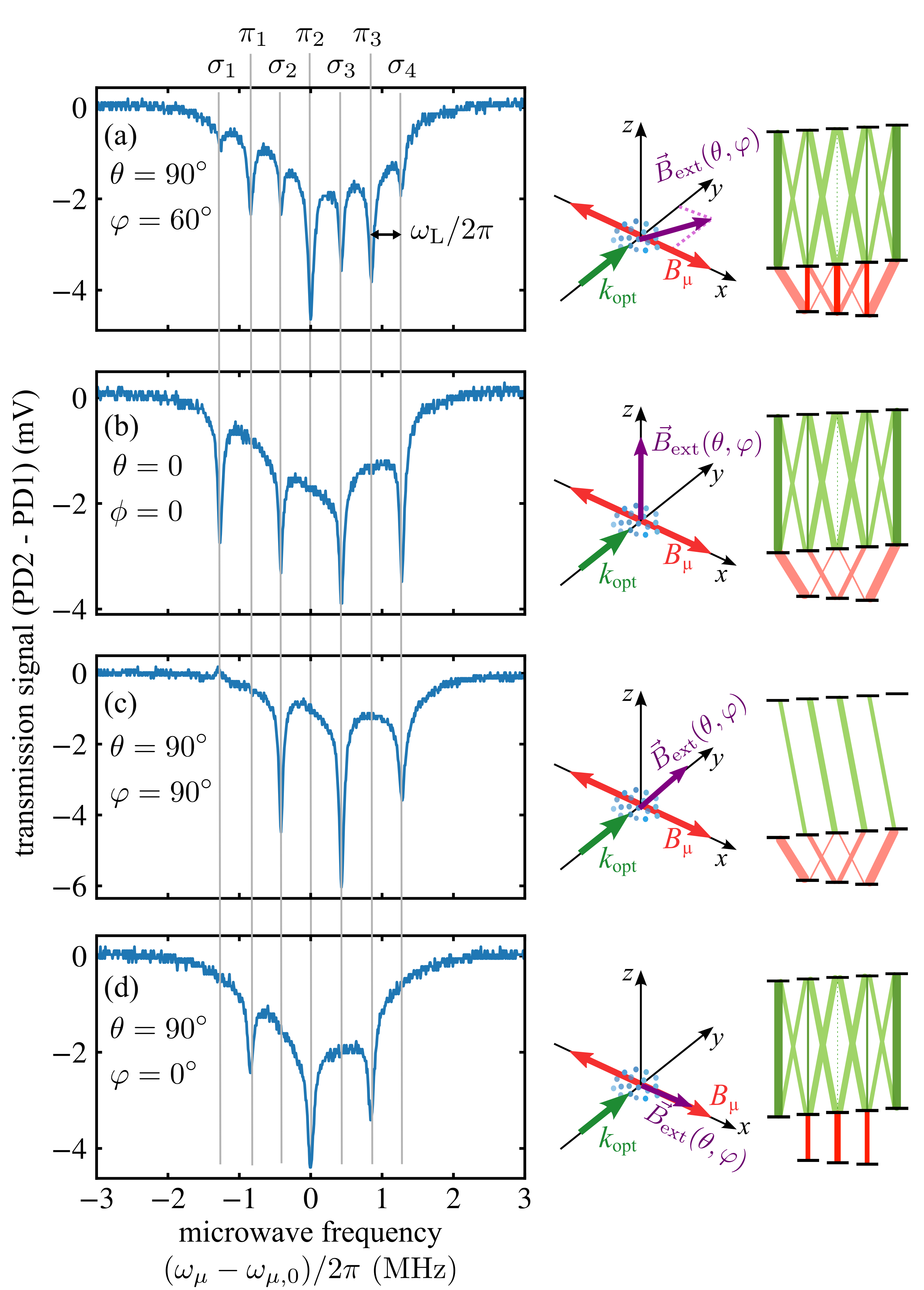}
    \caption{Double-resonance (DR) optical transmission signals, as a function of microwave detuning, for four different $\vec{B}_{\text{ext}}(\theta,\varphi)$ orientations. The microwave transitions are numbered and indicated at the top, as in Fig.~\ref{fig:intro}, noting that $\sigma_{2,3}$ represent two degenerate transitions. (a) $\vec{B}_{\text{ext}}(90^\circ,60^\circ)$, all transitions are allowed; (b) $\vec{B}_{\text{ext}}(0^\circ,0^\circ)$ (along $\hat{z}$, $\phi$ is arbitrary), $\pi$-polarized microwave transitions are forbidden; (c) $\vec{B}_{\text{ext}}(90^\circ,90^\circ)$ (along $\hat{y}$, parallel to optical propagation), $\pi$-polarized microwave transitions and $\pi$- and $\sigma^+$-polarized optical transitions are forbidden; (d) $\vec{B}_{\text{ext}}(90^\circ,0^\circ)$ (along $\hat{x}$, parallel to microwave magnetic field), $\sigma^\pm$-polarized microwave transitions are forbidden.
    In the level diagrams to the right, light and dark   lines depict the allowed $\sigma$ and $\pi$ microwave (lower, red) and optical (upper, green) transitions; the width of these lines are proportional to the transition amplitudes-squared.}\label{fig:signals}
\end{figure}

The amplitude of the external magnetic field is found using a measure of the Zeeman splitting in the system. For measurements of the optical absorption through the system,  changes are observed when the microwave frequency is resonant between two Zeeman sublevels: due to microwave-assisted optical pumping~\cite{Tretiakov2024}, the absorption increases. Sweeping the microwave frequency across $\omega_{12}$ reveals seven DR features, as shown in Fig.~\ref{fig:signals}. These features correspond to nine microwave transitions, of which two pairs ($\sigma_2$ and $\sigma_3$ in Fig.~\ref{fig:intro}a) overlap in frequency. This results in  seven DR features, as seen in Fig.~\ref{fig:signals}a. 
The Larmor frequency $\omega_L$ separates adjacent DR features in the absorption trace (Fig.~\ref{fig:signals}a), and depends on the total amplitude of the external magnetic field, $|\vec{B}_{\text{ext}}|$, through $\omega_{\rm L}/2\pi = \lambda_g |\vec{B}_{\text{ext}}|$, where $\lambda_g$ is the gyromagnetic ratio ($\lambda_g = 7$ kHz/$\mu$T for the ground state).

The amplitudes of these DR features also encapsulate vectorial information, that is, the orientation of  $\vec{B}_{\text{ext}}$ with respect to the optical propagation vector $\vec{k}_{\rm opt}$ and the microwave magnetic field polarization $\vec{B}_\mu$. 
The directionality of the field can be accessed because the selection rules that govern atomic transitions mean that the polarization of the electromagnetic fields driving transitions \emph{relative to} a quantizing direction dictate which transitions are possible and, together with the transition strengths, how strong they are. 
The different strengths (and thus Rabi frequencies) of these transitions mean that the direction of the external field changes the absorption properties of the atomic transitions. In the cases studied here, the static external field $\vec{B}_{\text{ext}}$ sets the quantization direction, and the coupling fields are decomposed into components labeled $q\in \{\sigma^+, \sigma^-, \pi\}$, which drive transitions between Zeeman sublevels $m_F$ and $m_F^\prime = \{m_F+1, m_F-1, m_F\}$, respectively.

Each component’s contribution depends on the quantization axis and the specific initial and final states involved. 
For the microwave magnetic field, for example, $\vec{B}_{\mu} = B_{\mu} \hat{{e}}_{\mu}$ can be  decomposed in terms of its polarization, with unit vector 
 \begin{align}
     \hat{{e}}_{\rm \mu}= b_{\sigma -} \left(\frac{\hat{x} - i \hat{y}}{\sqrt{2}}\right) + b_{\pi} {\hat{z}} + b_{\sigma+} \left(\frac{\hat{x} + i \hat{y}}{\sqrt{2}}\right),
     \label{eq:mwpol}
 \end{align} 
where $b_q$ are amplitudes of each polarization component. Assuming the external field is along $\hat{z}$, the Rabi frequency matrix element depends on $\vec{\mu}\cdot\hat{{e}}_{\mu} = [\mu_+b_{\sigma +} + \mu_-b_{\sigma-} + \mu_z b_\pi]$, i.e.,  $\Omega_{\mu}^{q} \propto b_{q} B_\mu \bra{F^*,m+q} \mu_q  \ket{F,m}/\hbar$, where the index $q = \{+,0,-\} = \{\sigma^+,\pi,\sigma^-\}$  denotes the polarization. A similar treatment can be applied to the optical field, whose polarization $\vec{E}_{\rm opt} = E_{0} \hat{{e}}_{\rm opt}$ is decomposed in the same basis as
\begin{align}
    \hat{{e}}_{\rm opt} =  a_{\sigma-} \left(\frac{\hat{x} - i \hat{y}}{\sqrt{2}}\right) + a_{\pi} {\hat{z}} + a_{\sigma+} \left(\frac{\hat{x} + i \hat{y}}{\sqrt{2}}\right),
    \label{eq:optpol}
\end{align}
where the amplitudes $a_q$ represent the amplitude of the optical polarization in each of three  configurations, and $\Omega_{\rm opt}^{q}  = a_{q}E_0 \bra{F',m+q}  d_{q}\ket{F,m}/\hbar$, where  $d_q \propto Y_{1,q}$, the spherical harmonic functions. 

As the direction of the external magnetic field [defined in polar coordinates as $\vec{B}_{\rm ext}(\theta,\varphi)$] rotates, the relevant coordinate system with respect to this quantization axis rotates, and the microwave and optical field polarizations must be defined with respect to $\tilde{\hat{z}} \parallel \vec{B}_{\rm ext}(\theta,\varphi)$.
It follows that the fields' polarization components will in turn depend on these angles via transformed amplitudes $\tilde{a}_q(\theta,\varphi)$ and $\tilde{b}_q(\theta,\varphi)$; consequently, the DR absorption amplitude becomes a function of \(\theta\) and \(\varphi\) (Appendix \ref{app:B}).

While precisely understanding the absorption feature amplitudes depends on a number of factors (both matters of control and environment), the premise of this technique is revealed when considering special cases. As an example, consider the geometry depicted in Fig.~\ref{fig:signals} and used in our experiments: the microwave magnetic field is along the cavity axis ($\hat{e_\mu} = \hat{x}$) and the optical propagation vector is along $\vec{k}_{\rm opt}\hat{y}$, with left circular polarization. When the external field $\vec{B}_{\rm ext}(\theta,\varphi)$ is aligned along $\hat{z}$  ($\theta = 0$, Fig.~\ref{fig:signals}b),  it is perpendicular to the microwave field, and only the $\sigma_\pm$ microwave transitions are allowed ($b_\pi = 0, b_{\sigma\pm} = 1/\sqrt{2}$), while the polarization is mixed for the optical field and all transition components are nonzero. In the measurement scheme used in this work, the optical absorption will change when the microwave frequency is tuned to one of the allowed resonances due microwave assisted optical pumping into the target state. For this geometry, only the four $\sigma^\pm$ microwave features are present, at $\omega_\mu-\omega_{\mu,0} = \pm \omega_{\rm L}, \pm 3\omega_{\rm L}$). 

A similar configuration arises when $\vec{B}_{\rm ext}(\theta = \pi/2,\, \varphi = \pi/2)$ is aligned along $\hat{y}$ (Fig.~\ref{fig:signals}c):  again, the microwave $\sigma_\pm$ transitions are allowed ($b_\pi = 0, b_{\sigma\pm} = 1/\sqrt{2}$), but now the circularly polarized optical field drives only the $sigma_-$ optical transitions ($a_\pi = a_{\sigma+}= 0, a_{\sigma-} = 1$), and the state $\ket{F = 2,\, m'_F = -2}$ becomes a dark state. The absorption traces shows features at the same microwave detunings as in Fig.~\ref{fig:signals}b, but the feature at $\omega_\mu-\omega_{\mu,0} = - 3\omega_{\rm L}$ is small and inverted, and all the peaks are of different magnitude due to different optical pumping processes in this configuration.

When the external field points towards along the $\hat{x}$-axis ($\theta = \pi/2,~\varphi = 0$) (Fig.~\ref{fig:signals}d),  the $\pi$ microwave transitions are now allowed and the $\sigma_\pm$ transitions are forbidden ($b_\pi = 1$, $b_{\sigma\pm} = 0$), and all polarizations of the optical field are present. Here, three features remain in the DR spectrum, now at $\omega_\mu-\omega_{\mu,0} = 0,\pm2\omega_{\rm L}$. For intermediate angles, all polarization amplitudes are non zero, and transitions can be found for all resonances, as seen in Fig.~\ref{fig:signals}a. In general, varying both $\theta$ and $\varphi$ changes field amplitudes, driving both the microwave and optical transitions, and results in distinct ground-state population distributions and therefore, distinct absorption spectra.

In practice, the warm-atom system presents a number of complicating factors when it comes to predicting the exact amplitudes of the absorption features.  In this system, the absorption is affected by many factors including: optical pumping from mixed polarizations~\cite{Tretiakov2024}; relaxation among the ground states due to thermal and collisional effects that redistribute populations in the ground states; microwave cavity linewidth; and off-resonant optical excitations~\footnote{In warm atom systems like this one, Doppler broadening populates velocity classes for which this laser is resonant to the $F'=1,2,3$ excited levels}. The complicated dynamics between the 24 ground and excited states would require a demanding numerical calculation that, despite the best first-principles modelling, will still require calibration inputs from the experiment. Instead, we extract the vector information of $\vec{B}_{\rm ext}(\theta, \varphi)$ using a machine learning approach that takes into account the richness of these spectra without needing prior information about the specific experimental conditions. 

\section{Magnetometer configuration} 

 At the heart of this magnetometer, as shown in Fig.~\ref{fig:intro}(b), is a $^{87}$Rb atomic vapor contained in a sealed $3\times 1 \times 1 ~\text{cm}^3$  rectangular quartz cell, centered within a copper cylindrical microwave cavity~\cite{Tretiakov2020-py,Ruether2021-rq,Smith2023}. This cavity has an internal diameter of 58~mm and a length of 51~mm. The dimensions of this cavity are chosen to resonate at $\omega_\mu/2\pi  = 6.834$~GHz in the TE$_{011}$ mode, addressing the magnetic dipole transition. This mode provides a nearly-uniform magnetic field throughout the vapor cell and is linearly polarized along the cavity axis $\hat{x}$ as shown in Fig.~\ref{fig:intro}(c). One cavity end cap position along $\hat{x}$ is adjustable, allowing precise tuning of cavity resonance and a 2~mm-wide annular groove machined into the cavity’s cap  lifts the frequency degeneracy between the TE$_{011}$ and TM$_{111}$ modes~\cite{Thal1979,Tretiakov2020-py}.   The cavity eliminates the need for an RF coil, making the magnetometer suitable for miniaturization, especially as the microwave cavity can be replaced by circuit-printed split-ring resonators~\cite{RoyChoudhury2016}.

To achieve a relatively wide cavity bandwidth (more than 2~MHz), sufficient to cover nearly all seven DR features at our desired external magnetic field, we operate the cavity in a low quality-factor (QF) regime.  A temperature controller maintains the cavity-cell setup at 30$^{\circ}$C, resulting in a QF of 2900~\cite{Smith2023}.
A relatively weak (100~$\mu$W) left-circularly polarized 780~nm laser is used to probe the population of the $F=2$ ground state through two small holes, each with a 1~mm radius, machined into the microwave cavity.
A photodiode (PD2) detects the DR signal on the output beam [Fig.~\ref{fig:intro}(b)]; this signal is recorded using an oscilloscope triggered by the microwave source. Another photodiode (PD1) measures the input beam intensity and is used as a reference signal to reduce common-mode noise.
The microwave source, connected to the cavity's coupler pin via an SMA cable, sweeps the frequency in a sawtooth ramp every 4 ms
around 6.834~682~GHz, traversing a span of 3~MHz.  Each trace is averaged 256 times to improve the signal-to-noise ratio. This produces a measurement time of 1.024~s at each field orientation.

Three sets of  Helmholtz coils, mounted orthoganally, generate the external magnetic field $\vec{B}_{\text{ext}}$, with currents adjusted to point the field in all directions. 
A computer synchronizes and automates the coils' current drivers and the oscilloscope to record large number of DR signal traces.

\section{Determining field orientation from DR spectra}
\begin{figure}
    \centering
    \includegraphics{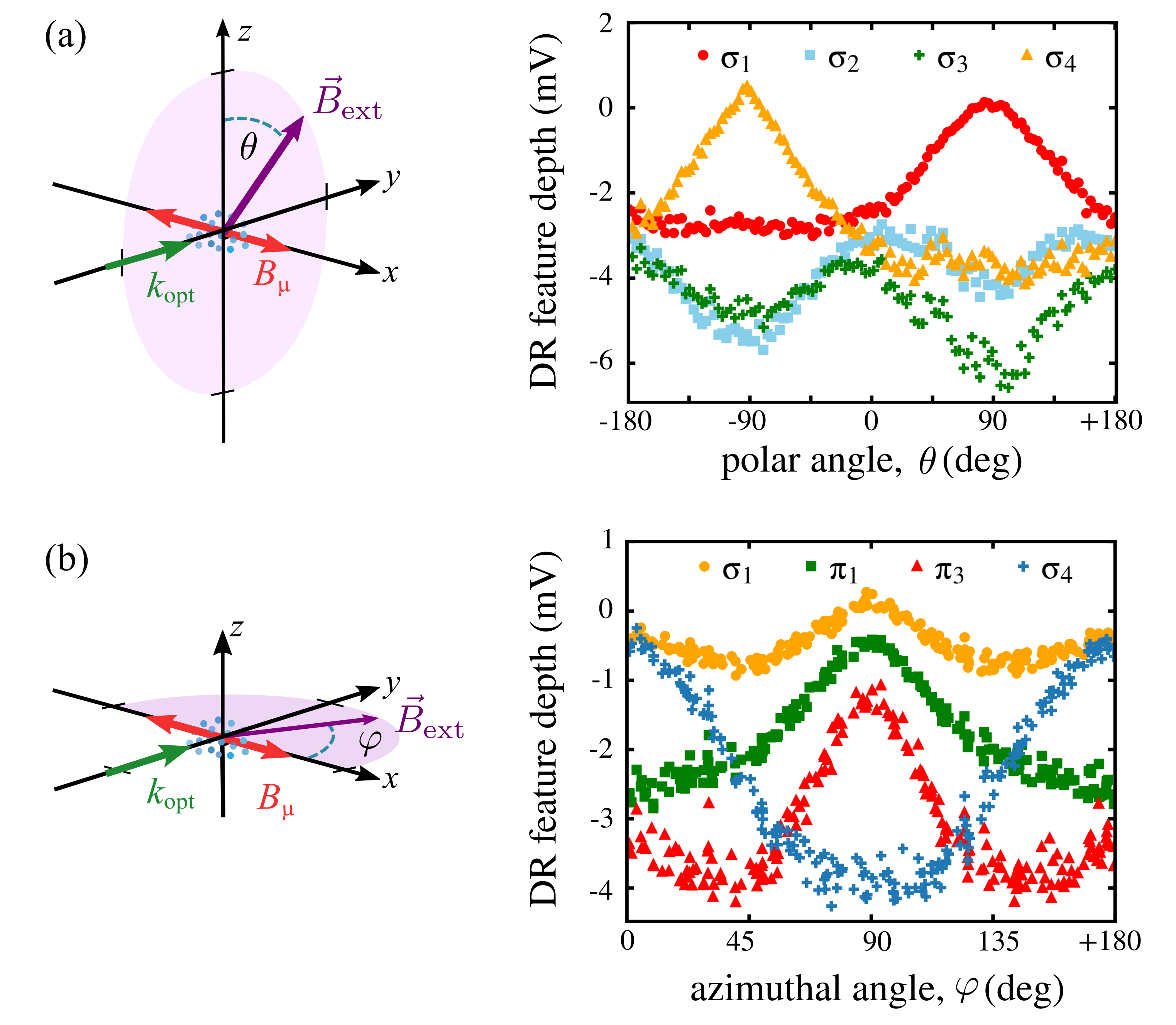}
    \caption{Variation in selected double-resonance (DR) transmission feature depth as a function of angles (a) $\theta$ in the $y$-$z$-plane ($\perp \vec{B}_\mu$), and (b) $\varphi$ in the $x$-$y$-plane ($\parallel \{\vec{B}_\mu, \vec{k}_{\rm opt}$\}), for LCP light propagating along the $\hat{y}$-axis and $\vec{B}_\mu$ along $\hat{x}$. In both cases, the features are labelled by the microwave transition (as numbered in Fig.~\ref{fig:intro}): (a) Depicts resonant microwave feature depths for transitions $\sigma_1$ (red circle), $\sigma_2$ (blue square), $\sigma_3$ (green cross), $\sigma_4$ (yellow triangle). (b) Depicts resonant microwave features depths for transitions $\sigma_1$ (yellow circle), $\pi_1$ (green square), $\pi_3$ (red triangle), $\sigma_4$ (blue cross). The magnitude of the external magnetic field $|\vec{B}_{\rm ext}| = 50~\mu$T is fixed, as in Fig.~\ref{fig:signals}.}
    \label{fig:ampvar}
\end{figure}

To calibrate this vector magnetometer, we measured the DR absorption spectra for a wide range of external field angles and found clear changes in the absorption features with changes to the field's orientation (Fig.~\ref{fig:signals}). 
To analyze these variations systematically, we consider two planes in which only one spatial angle varies. In the $\hat{y}$-$\hat{z}$ plane ($\varphi = \pm \pi/2$), only the six $\sigma$ microwave transitions, at four distinct frequencies, are allowed. The  amplitudes of the $\sigma$ microwave features vary with changes in the  angle $\theta$, as shown in Fig.~\ref{fig:ampvar}a.

\begin{figure*}[tb!]
    \centering
    \includegraphics{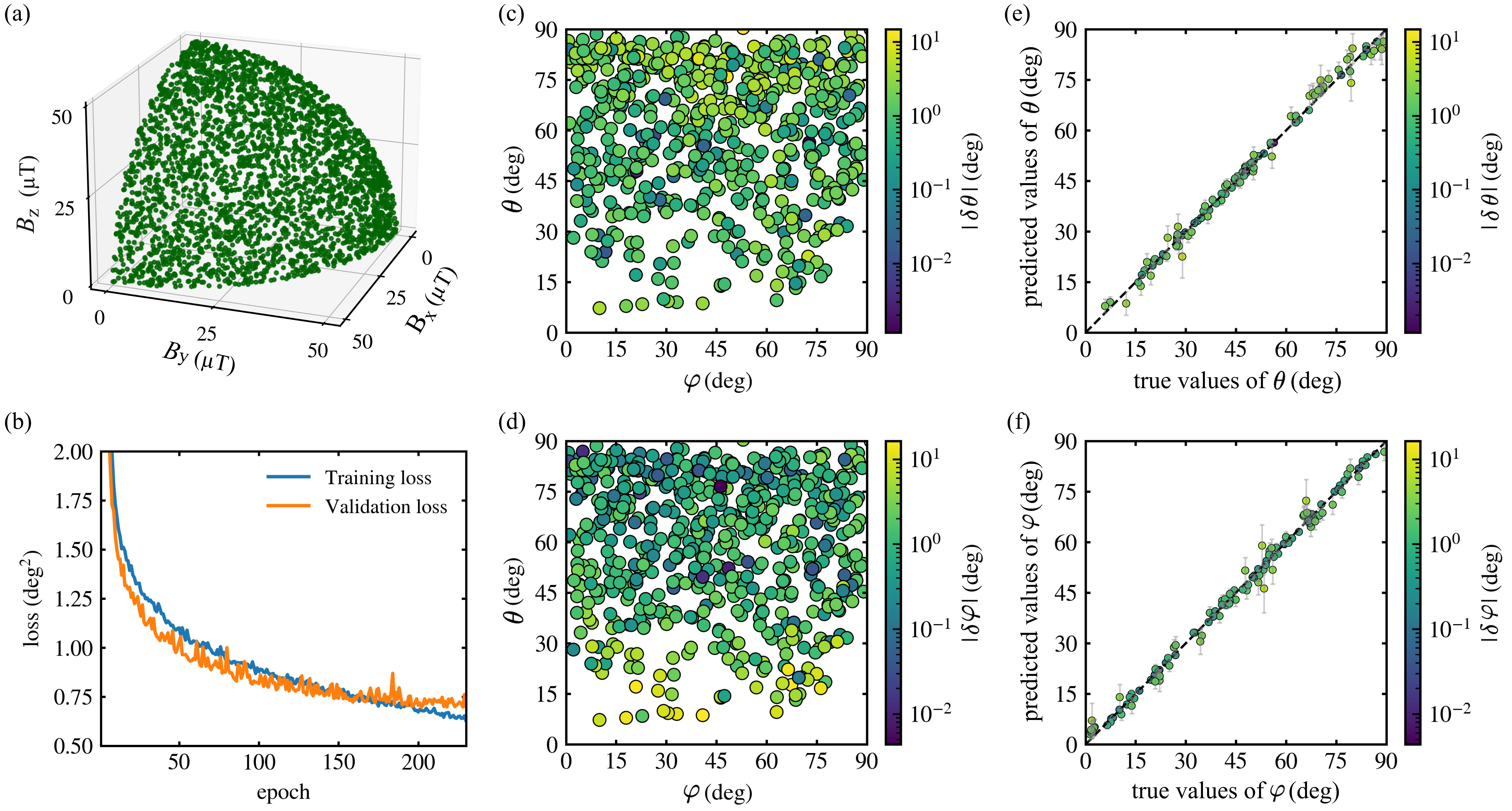}
    \caption{Machine-learning analysis of an atomic vector magnetometer. (a) Experiments were performed for 3000 randomly oriented $\vec{B}_{\rm ext}$ vectors (fixed magnitude of 50~$\mu$T) across the positive octant  in 3D Cartesian space. (b) As a function of epoch,  training (orange) and validation (blue) loss curves indicate ML model performance, with validation loss converging at 230 epochs. (c-d) For a validation set of 300 unseen datasets, the model's absolute prediction errors $|\Delta\theta|$ (c) and $|\Delta\varphi|$ (d) are shown mapped onto the $\theta$--$\varphi$ distributions. The color scale represents the magnitude of the angular error in degrees. (e-f) Predicted versus true values of $\theta$ (e) and $\varphi$ (f). Each point represents the mean prediction from an ensemble of five 1D-CNN models (bagging), with error bars and colourscale both indicating the standard deviation across the predictions. The dashed line (unit slope) represents the  prediction of a perfect model.}
    \label{fig:result1}
\end{figure*}

Figure~\ref{fig:ampvar}b shows a complementary scenario for a sweep of $B_{\rm ext}$ in the $\hat{y}$-$\hat{z}$ plane ($\theta = \pi/2$). As the angle $\varphi$ varies from $0$ to $\pi/2$, the $\pi$ microwave features diminish and eventually vanish while the the $\sigma$ microwave features gradually emerge.
These two  configurations demonstrate that well defined relationships exist among the DR features' amplitudes, and the information is encoded across all seven of the absorption features.
However, analytically modelling the full angular dependence in three-dimensional space introduces significant complexity, and fitting these data to realistic models soon lends itself to highly parameterized system that depends on physical processes including collisions between atoms and the container, the microwave cavity linewidth, optical pumping through off-resonant excited states, and other relaxation processes (such as transit-time broadening) that depend on the system temperature and size. Rather than attempt a fitting routine with a large number of parameters, many of which would need to be assumed, we introduce a machine learning model to analyze the variations in the spectral feature amplitudes with respect to spatial angle, and use the model to determine the orientation of the external magnetic field.

In comparing machine-learning approaches, we find that a one-dimensional convolutional neural network (1D-CNN) demonstrates the best performance, as it is well-suited to sequential input data and where nearby elements are closely related or exhibit a pattern~\cite{Goodfellow2016}. (The 1D-CNN shows excellent performance across applications, including time-series data, text classification, and DNA/RNA sequence analysis \cite{Zeng2016, Kim2023, Kim2014}.)
A 1D-CNN is built using TensorFlow/Keras~\cite{tensorflow2015-whitepaper, kerashub2024} and consists of three convolution pooling stages (Conv1D) with $32\to64\to128$ filters and ReLU activations, followed by max-pooling and dropout layers for feature extraction and regularization~\cite{Mehta2019}. After extracting features from the input sequences, the feature maps are then flattened and passed through a deep stack of ReLU-activated dense layers. The fully connected layers refine the extracted features before making predictions. The model is optimized with the Adam optimizer \cite{Mehta2019} and the Huber loss function~\cite{Hastie2009}. The Huber loss function was selected to mitigate the influence of outliers, which arose from a small number of corrupted or noisy traces in the dataset. This function is quadratic below the clip parameter, $\delta = 5^{\circ}$ (absolute error), and linear beyond it.

To train the CNN model, 3000 data points were randomly selected on the surface of one-eighth of a sphere with a 50~$\mu$T radius, as shown in Fig.~\ref{fig:result1}a. The model maps each spatial angle of $\vec{B}_{\text{ext}}(\theta,\varphi)$ vector to an 800-point absorption spectrum.
The architecture of the CNN model is designed  to avoid overfitting, which occurs when the model starts memorizing the training data and loses its ability to perform well on unseen (test) datasets. Figure~\ref{fig:result1}b shows the loss on the training and validation datasets during model training. The validation set comprises a random 10$\%$ of the entire dataset. An early-stopping criterion was applied to terminate training after 30 consecutive epochs without improvement. The model converged within approximately 230 epochs to 0.793~$\text{deg}^2$ error on the unseen (test) datasets. To improve accuracy, we employed bagging \cite{Mehta2019} with five different 1D-CNN models, each configured with varying numbers of neurons in the hidden layers. 

Figures~\ref{fig:result1}c-f illustrate the performance of the machine learning model in predicting the two spatial angles, $\theta$ and $\varphi$. The absolute deviations of the predicted values from the true values are represented by the color scale. Figures~\ref{fig:result1}c and d show the angular errors $|\delta\theta|$ and $|\delta\varphi|$ across the $\theta$--$\varphi$ distribution, while Figs.~\ref{fig:result1}e and f display the predicted versus true values for $\theta$ and $\varphi$, respectively. The dashed line (unit slope) indicates the ideal case of perfect prediction, where predicted values are equal to true values. The mean errors of the two spatial angles over a test-dataset of 300 samples are approximately $\delta\theta \approx 1.3^\circ$ and $\delta\varphi \approx 1.7^\circ$. The minimum deviations observed are $\delta\theta_{\text{min}} \approx (1.1 \times 10^{-3})^\circ$ at $\theta=47^\circ,\varphi=19^\circ$ and $\delta\phi_{\text{min}} \approx (4.3 \times 10^{-3})^\circ$ at $\theta=76^\circ,\varphi=45^\circ$. However, in certain regions, weak predictions result in an increased overall standard deviation. This degradation is particularly evident near the coordinate axes, where the prediction quality diminishes, as reflected by the reduced amplitudes of the $\pi$ and $\sigma$ peaks, approaching the noise floor. Such effects could be mitigated through enhanced system stabilization.
The primary source of deviation arises from uncertainty in the orientation of the magnetic field generated by the Helmholtz coils. Although precise calibration is performed for each coil pair by using the linear relationship between coil currents and the measured Larmor frequency ($\omega_L$) from the DR signal (one of the main advantages of atomic magnetometry being this self-calibration), imperfections in the orthogonality of the three coil sets and minor current drifts may result in discrepancies between the actual and intended vector fields, contributing to increased deviation in the measured angles from the expected ones~\cite{Meng2023}.

\begin{figure}
    \centering
    \includegraphics{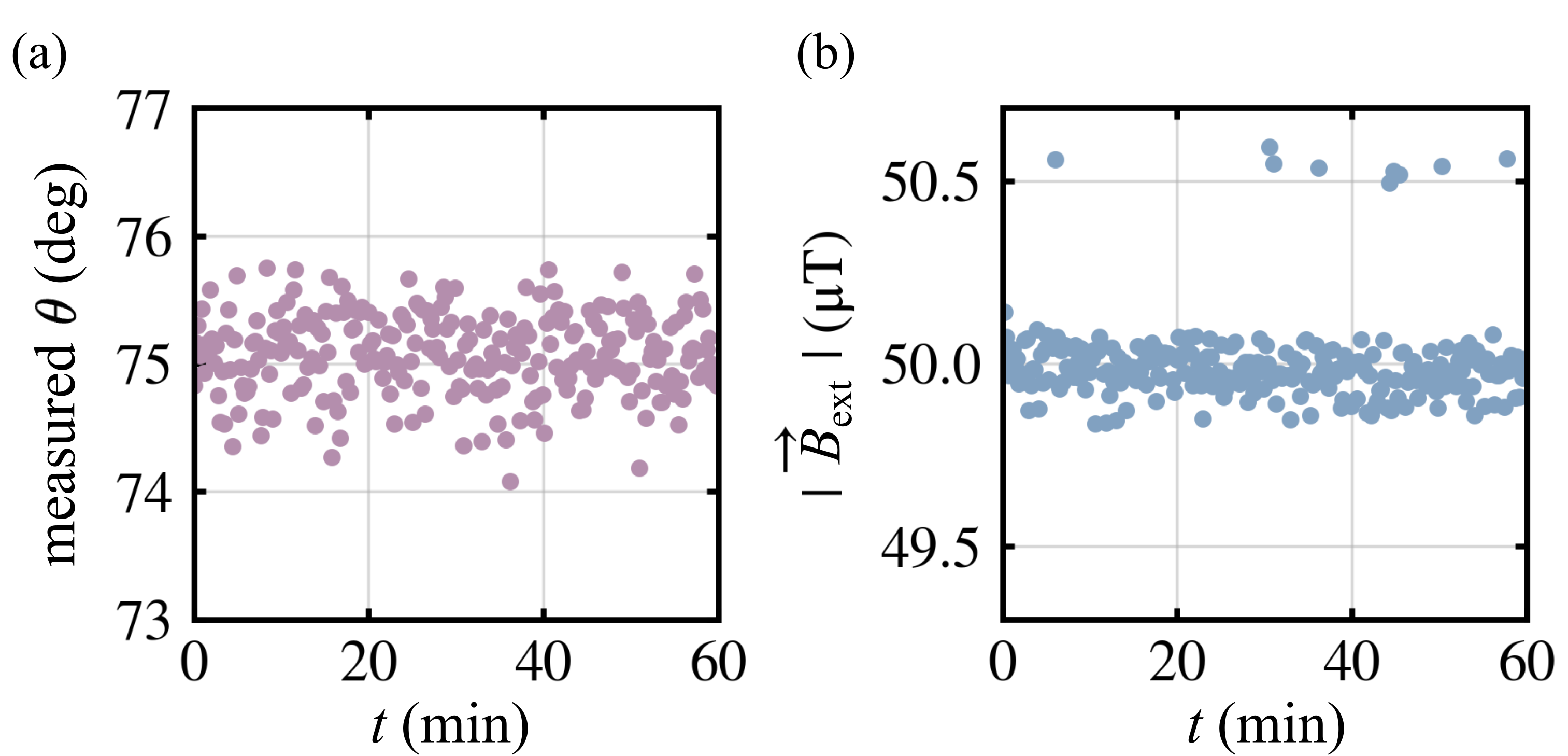}
    \caption{Fluctuations of the applied magnetic field $|B_{\rm ext}|$, amplitude of 50~$\mu$T at $\varphi=90$ and $\theta = 75^\circ$, over 60 minutes in (a) magnetic field orientation and (b) amplitude, as determined by our magnetometer and ML model. Under our unshielded experimental conditions, including background field and instrument (current supply) drift/noise, our measurement precision is limited to the standard deviations from these measurements:  $0.4^\circ$ angular and 115~nT amplitude.}
    \label{fig:fluc}
\end{figure}

Given that this magnetometer operates in an unshielded environment, natural variations in the Earth’s magnetic field throughout the day can further shift the true field values. Additional small environmental fluctuations, such as changes in temperature, coil currents, optical input power and polarization also influence the system.
These fluctuations are illustrated in Fig.~\ref{fig:fluc}, which shows variations in the measured angle and amplitude under constant coil current conditions over a one-hour period.
For an external magnetic field of $|\vec{B}_{\text{ext}}(75^\circ, 90^\circ)| = 50~\mu\text{T}$, the standard deviation in the measured angle is approximately $0.4^\circ$, while for the amplitude it is 115~\text{nT}. To overcome these limitations, a calibrated sensor would need to be used to perform comparison measurements with our system. Additionally, we note that the minimum measurable magnetic field amplitude for this system is around 10~$\mu$T, limited--not by the noise floor--but by the linewidth of the DR absorption features of about 80~kHz.

\section{Conclusion}
In this work, we present a three-axis vector magnetometer based on the double-resonance absorption in $^{87}$Rb. Using a 1D-CNN machine learning model, we extract magnetic field vector and magnitude information from the DR spectrum, obtaining vector information from the spectral feature amplitudes, with precision of 1.3$^\circ$ for the polar angle and 1.7$^\circ$ for the azimuthal angle; we obtain the magnetic field amplitude from the spacing between the DR features, producing a sensitivity of 115~nT/$\sqrt{\text{Hz}}$ (see Appendix \ref{app:C}) for an external field of 50 $\mu$T. Fluctuations in the magnetic field amplitude are reflected in the horizontal DR absorption dips displacement, not limited by the vertical noise on the DR signal.

The primary factors limiting measurements of sensitivity and precision in this proof-of-concept work are fluctuations in optical power, noise and thermal drift in the Helmholtz coil currents generating the calibrating fields, and variations in the background magnetic fields over the course of calibration runs. Advanced implementations of this scheme, including dual-device reference calibration and ML training, as well as enhanced optical power stabilization, will provide significant improvement in both the sensitivity and precision of this approach, allowing for measurements at the fundamental limits.

The unshielded device presented in this work is tailored for Earth-scale magnetic fields and operates using  a single optical beam and a microwave cavity. Importantly, it does not require external polarization modulation of the optical field, mechanical rotation, or multiple optical beams. These characteristics, combined with the self-calibration capacity and the ability to operate at room temperature, underscore this magnetometer’s strong potential for commercialization as a miniaturized, portable device suitable for a wide range of applications.

\begin{acknowledgments}
We would like to acknowledge Y. Y. Lu for  preliminary measurements, J. Shi for providing assistance with the data acquisition,  and Dr. C. A. Potts for machining the microwave cavity. This work is supported
by funding sources NSERC (RGPIN-2021-02884 and ALLRP 578468-22), DND IDEaS MN4-056, Alberta Innovates, Canada Foundation for Innovation (36130), the Province of Alberta, and the Canada Research Chairs (CRC) Program. As researchers at the University of Alberta, we acknowledge that this work was conducted on Treaty 6 territory, and that we respect the histories, languages, and cultures of First Nations, Métis, Inuit, and all First Peoples of Canada, whose presence continues to enrich our vibrant community.
\end{acknowledgments}

\appendix
\section{Master equation solution}\label{app:A}
To find the optical absorption of the microwave-optical double resonance in the three-level system described in Sec.~\ref{DR absorption}, we first solve the Lindblad master equation for the given Hamiltonian. By doing so, we can derive the optical Bloch equations~\cite{Bonifacio1965} and determine the population dynamics of each state.
This master equation \cite{Breuer2007} is
\begin{equation}
\dot{\rho} = -\frac{i}{\hbar}[H, \rho] + \sum_i \gamma_i \left( L_i \rho L_i^\dagger - \frac{1}{2} \left\{ L_i^\dagger L_i, \rho \right\} \right),
\end{equation}
where the second term is 
\begin{equation}
\begin{pmatrix}
\gamma (\rho_{22} - \rho_{11}) + \frac{\Gamma}{2} \rho_{ee} & -\gamma \rho_{12} & -\Gamma' \rho_{1e} \\
-\gamma \rho_{12}^{*} & \gamma (\rho_{11} - \rho_{22}) + \frac{\Gamma}{2} \rho_{ee} & -\Gamma' \rho_{2e} \\
-\Gamma' \rho_{1e}^{*} & -\Gamma' \rho_{2e}^{*} & -\Gamma \rho_{ee}
\end{pmatrix},
\end{equation}
and $\rho_{ij}$ are the density matrix components $\{i,j\}\in \{1, 2, e\}\ $, and $\Gamma$ and $\gamma$ are the optical and microwave relaxation rates, respectively. For the experimental conditions in this work, the orders of magnitude are such that $\gamma \sim 2\pi \times 100 ~\text{Hz} \ll \Gamma \sim 2\pi \times 6~\text{MHz}$, and we can assume $\Gamma' ={(\Gamma+\gamma)}/{2} \approx{\Gamma}/{2}$. 

We use the Beer-Lambert law~\cite{Foot2005} to calculate the absorption of optical light after it propagates a short distance \( L \) (the width of the vapor cell) through a dilute gas. The intensity of this light at frequency \( \omega_p \) is
\begin{equation}
I(\omega_p, L) = I_0 \left[ 1 - \sigma(\omega_p) L {N} (\rho_{22} - \rho_{ee}) \right],
\end{equation}
where \( I_0 \) is the intensity before entering the vapor cell, \( \sigma(\omega_p) \) is the absorption cross-section, and \( (\rho_{22} - \rho_{ee}) \propto (\Omega_p, \Omega_\mu) \) are obtained by solving the Bloch equations in the steady state. The optical absorption amount, $\Delta I$, is proportional to the Rabi frequencies of both the microwave and optical fields, $\Omega_\mu$ and $\Omega_p$, respectively. These Rabi frequencies depend on the amplitudes of the applied fields and the dipole matrix elements associated with the initial and final atomic states.
An external magnetic field lifts the degeneracy in the atomic hyperfine manifold via the Zeeman effect. This results in the splitting of energy levels into Zeeman sublevels, with frequency separations determined by the Larmor frequency, $\omega_L$. Consequently, the transition amplitudes and thus the coupling strengths between individual Zeeman sublevels vary, depending on the polarization and relative orientation of the applied fields.

In a realistic atomic system, particularly one involving a three-level configuration, these Zeeman sublevels are not isolated. The population dynamics and transition probabilities are influenced by nearby sublevels, making it inappropriate to treat them as independent. Therefore, a complete and accurate description of the interaction must include the full Zeeman structure.
However, to illustrate and analytically support the dependence of optical absorption on the Rabi frequencies, a simplified three-level model can serve as a useful approximation. This model captures the essential physics of the interaction while acknowledging that a more comprehensive model would incorporate all relevant Zeeman sublevels.

\section{Dependence of Rabi frequencies on external field orientation}\label{app:B}
To understand the angular dependence of the Rabi frequencies as discussed in Sec.~\ref{Absorption dependence on external magnetic field}, we examine the radiation's polarization variation with respect to the quantization axis' orientation, $\parallel \vec{B}_{\rm ext}(\theta,\varphi)$.
For an electromagnetic field $\vec{\mathcal{F}}(\theta,\varphi) = \mathcal{F}_{-1}\hat{e}_{-1} + \mathcal{F}_{0}\hat{e}_{0} + \mathcal{F}_{+1}\hat{e}_{+1} $ where
\begin{align}
    \hat{e}_{\pm 1} = \left(\frac{\hat{x} \mp i \hat{y}}{\sqrt{2}}\right), \qquad{\rm and} \qquad \hat{e}_{0} = \hat{z}.    
\end{align}
In a rotation of the quantization axis from $\hat{z}$ to $(\theta,\varphi)$ (in the rank-1 spherical basis)~\cite{auzinsh2010optically}, the component amplitudes $\mathcal{F}_q \in (a_q,b_q)$  become
\begin{align}
\tilde{\mathcal{F}}_q(\theta,\varphi)
=
\sum_{q'=-1}^{+1} e^{-iq'\varphi} \mathcal{R}^{(1)}_{q,q'}(\theta,\varphi) \mathcal{F}_{q'},
\end{align}
where 
\begin{align}
\bm{\mathcal{R}}^{(1)}(\theta,\varphi) =
\begin{pmatrix}
\cos^2\frac{\theta}{2}
& -\frac{1}{\sqrt{2}}\sin\theta
& \sin^2\frac{\theta}{2}
\\
\frac{1}{\sqrt{2}}\sin\theta
& \cos\theta
& -\frac{1}{\sqrt{2}}\sin\theta
\\
\sin^2\frac{\theta}{2}
& \frac{1}{\sqrt{2}}\sin\theta
& \cos^2\frac{\theta}{2}
\end{pmatrix},
\end{align}
is the rotation matrix.

\subsection{Microwave magnetic field}
In the experimental configuration used in this work, the microwave field is 
\begin{align}
    \vec{B}_{\mu} = B_{\mu}~\hat{x} = \frac{B_{\mu}}{\sqrt{2}}\left(\hat{e}_{-1}-\hat{e}_{+1}\right),
\end{align}
meaning that in the notation used in the main text, $b_{\sigma-} = - b_{\sigma+} = {|B_{\mu}|}/{\sqrt{2}}$ and $b_{\pi} = 0$.
Under a rotation of the external quantizing field (labeled with tilde), the field in the transformed basis is $ \vec{\tilde{B}}_{\mu} = \sum_q \tilde{b}_{q}\hat{\tilde{e}}_q$, and the three polarization components are
\begin{align}
   \tilde{b}_{\sigma\pm}(\theta,\varphi)  &=\frac{1}{\sqrt{2}} \left[ \pm \cos\varphi \cos\theta + i \sin \varphi\right],
   \\
     \tilde{b}_{\pi}(\theta,\varphi)  &= \sin\theta\cos\phi,
\end{align}
resulting in an effective Rabi frequency that varies with orientation. Recalling the expression from the main text, the Rabi frequency assigned to a particular transition is now angle-dependent: $\Omega_{\mu}^{q}(\theta,\varphi) \propto \tilde{b}_{q}(\theta,\varphi) B_{\mu} \bra{F^*,m+q} \mu_q  \ket{F,m}/\hbar$,
where the transition amplitudes are a property of the particular levels involved, and are summarized in Table~\ref{tab:magdipoles}.

\begin{table}[tb!]
\centering
\label{tab:magdipoles}
\caption{Magnetic-dipole transition amplitude $\bra{F'=2,m_F'} \mu_s \ket{F=1, m_F}$.~\cite{ibali2017} }
\begin{tabular}{lccccc}
\toprule
\toprule
\textbf{ } & $\left\langle 2,-2\right |$ & $\left\langle 2,-1 \right |$ & $\left\langle 2,0 \right |$ & $\left\langle 2, +1 \right|$ & $\left\langle 2, +2 \right|$ \\
\midrule
$\left |1, -1 \right\rangle$ & $-\sqrt{3/2}$   & $\sqrt{3}/{2}$ & $-{1}/{2}$ & 0& 0\\
$\left |1, 0 \right\rangle$ & 0 & $-{\sqrt{3}}/{2}$ & $1$ & $-\sqrt{3}/{2}$  & 0\\
$\left |1, +1 \right\rangle$ & 0& 0& $-{1}/{2}$ & $\sqrt{3}/{2}$ & $-\sqrt{{3}/{2}}$ \\
\bottomrule
\bottomrule
\end{tabular}
\end{table}

In the protocol used in this work, the amplitudes of the features associated with each microwave transition depend on the orientation of the external magnetic field through the Rabi frequency dependence, which gives the strength of that particular transition.  While this simple analysis does not take into account collisional or thermal redistributions of the atoms in a vapour that also affect the steady-state populations and transition amplitudes, it provides a framework for understanding the origin of the signal's angular dependence in this system.

\subsection{Optical electric field}
The optical field involved in the electric dipole transitions will undergo a similar rotation to the microwave magnetic field. In the experimental configuration used in this work, the optical field is left-circularly polarized and travelling down the $\hat{y}$-axis. In this case, the optical electric field is 
\begin{align}
    \vec{E}_{\rm LCP} =  E_{0}  \left(\frac{\hat{x}-i\hat{z}}{\sqrt{2}}\right),
\end{align}
such that $a_{\sigma-}=-a_{\sigma+} = 1/2$ and $a_{\pi} = i/\sqrt{2}$
The representation of the field $ \vec{\tilde{E}}_{\rm opt} = E_0\sum_q \tilde{a}_{q}\hat{\tilde{e}}_q$, in terms of the rotated components 
\begin{align}
    \tilde{a}_{\sigma\pm}(\theta,\varphi)&=\left[-\frac{1}{2}\cos\theta\cos\varphi \pm \frac{i}{2}\left(\sin\theta \pm \sin\varphi\right)\right]
    \end{align}
\begin{align}
   \tilde{a}_{\pi}(\theta,\varphi) &= \frac{1}{\sqrt{2}} \left[ \sin\theta\cos\varphi + i\cos\theta \right],
\end{align}
provides a Rabi frequency for a particular transition that is  angle-dependent: $\Omega_{\rm opt}^{q}(\theta,\varphi) \propto \tilde{a}_{q}(\theta,\varphi) E_0 \bra{F{''},m+q} d_q  \ket{F',m}/\hbar$.

In the ambient-temperature systems used in this work, the dependence on these state-dependent Rabi frequencies is more complicated for the optical field than it is for the microwave field, due to the fact that the Doppler broadening makes all excited state hyperfine levels accessible, and so many transitions must be included to fully account for the dynamics in this system. Through a complex optical pumping process, a steady state will be reached in each angular configuration, but the details of this will depend not only on geometry, but also on the nonradiatie relaxations processes involved in this system. It is for this reason that we turn to a machine-learning analysis in this work, which can analyse the data for the angular dependence without relying on analytic or numerical modelling of every physical process governing the dynamics.

\section{Magnetometer sensitivity}\label{app:C}
Unlike conventional magnetometers, whose sensitivity is usually expressed as the noise‑equivalent magnetic field (NEMF) derived from the zero‑field power‑spectral density (PSD), i.e.\ the smallest \emph{vertical} field fluctuation that rises above the noise floor—our instrument is limited instead by the \emph{horizontal} (time‑axis) resolution of its time‑domain trace. The minimum detectable signal is therefore governed by the spacing between successive data points rather than by the noise level at zero field. For this reason we state the sensitivity of the device as the smallest magnetic‑field increment that can be resolved after one second of averaging, which can be calculated as \cite{Kiehl2025,Zhang2021}
\begin{align}
  S_{B} = |\delta B|\sqrt{t_{m}},
\end{align}
where $S_{B}$ is the sensitivity, $t_m$ the measurement time, and $\delta B$ the minimum resolvable magnetic field amplitude deviation. This equation can also be used for calculating the angular sensitivity. 


%

\end{document}